\documentclass{svproc}

\usepackage{url}

\usepackage{caption}
\usepackage{subcaption}
\usepackage{cite}
\usepackage{graphicx}
\usepackage{textcomp}
\usepackage{xcolor}
\usepackage{algorithm}
\usepackage{tabularx}
\usepackage{adjustbox}
\usepackage{float}
\usepackage{placeins}
\usepackage{booktabs}

\begin{document}
	\mainmatter              % start of a contribution
	
	% Political Power Structure behind Smart Sanctions
	\title{Identifying the Hierarchical Influence Structure Behind Smart Sanctions Using Network Analysis}
	\titlerunning{Hierarchical Influence Structure Behind Sanctions}  % abbreviated title (for running head)
	%                                     also used for the TOC unless
	%                                     \toctitle is used
	%
	\author{Ryohei Hisano\inst{1} \and Hiroshi Iyetomi\inst{2} \and
		Takayuki Mizuno\inst{3}}
	\authorrunning{Hisano et al.} % abbreviated author list (for running head)
	%
	%%%% list of authors for the TOC (use if author list has to be modified)
	%\tocauthor{Ivar Ekeland, Roger Temam, Jeffrey Dean, David Grove,
	%Craig Chambers, Kim B. Bruce, and Elisa Bertino}
	%
	\institute{The University of Tokyo\\
		\and
		Niigata University,
		\and
		National Institute of Informatics
		%Laboratoire d'Analyse Num\'{e}rique, B\^{a}timent 425,\\
		%F-91405 Orsay Cedex, France
		%\and
		%Universit\'{e} de Paris-Sud,
		%Laboratoire d'Analyse Num\'{e}rique, B\^{a}timent 425,\\
		%F-91405 Orsay Cedex, France
	}

	\maketitle              % typeset the title of the contribution
	
	\begin{abstract}
		
		Smart sanctions are an increasingly popular tool in foreign policy. Countries and international institutions worldwide issue such lists to sanction targeted entities through financial asset freezing, embargoes, and travel restrictions. The relationships between the issuer and the targeted entities in such lists reflect what kind of entities the issuer intends to be against. Thus, analyzing the similarities of sets of targeted entities created by several issuers might pave the way toward understanding the foreign political power structure that influences institutions to take similar actions.  In the current paper, by analyzing the smart sanctions lists issued by major countries and international institutions worldwide (a total of 73 countries, 12 international organizations, and 1,700 lists), we identify the hierarchical structure of influence among these institutions that encourages them to take such actions. The Helmholtz--Hodge decomposition is a method that decomposes network flow into a hierarchical gradient component and a loop component and is especially suited for this task. Hence, by performing a Helmholtz--Hodge decomposition of the influence network of these institutions, as constructed from the smart sanctions lists they have issued, we show that meaningful insights about the hierarchical influence structure behind smart sanctions can be obtained.
		
		%Specifically, we derived three interesting observations from our Helmholtz--Hodge analysis. First, we found that for smart sanctions lists against Iran and North Korea, while the United Kingdom, United Nations, and the United States are at the top of the hierarchy of influencing other major countries and international institutions, the United Nations Security Council is clearly at the bottom of the hierarchy, which indicates its complex approval mechanism.  Second, for the smart sanctions against Libya, the United Nations, ICTR, United States, and Japan are the key influencers in the upper stream of the hierarchy, and INTERPOL and the European Union are located in the middle of the hierarchy, acting as hubs aggregating information. Another intriguing example is the smart sanctions concerning financial crimes. In this example, there are four different communities (North America, Europe, island nations, and other countries), and each community focuses on different sets of targeted entities.
		
		%Furthermore, for some of the smart sanctions list categories, the influence network is governed almost entirely by a hierarchical structure. Still, for others, there is a significant amount of loop flow, which indicates that countries and international institutions actively influence each other. Our analysis enables a better understanding of the hidden global political structure behind smart sanctions.

		% We would like to encourage you to list your keywords within
		% the abstract section using the \keywords{...} command.
	\end{abstract}		
	\section{Introduction}

	As globalization progresses, economic trade among nations has been growing at an increasing rate~\cite{Steger2017}. There is no doubt that this global expansion of economic trade contributes to the growth of prosperity in the global economy. Still, the rising geographical and cultural distances among the participants make it challenging to avoid trading with entities (such as companies and individuals) that are involved with illegal criminal activities such as money laundering, terrorism, drug cartels, and human trafficking.  To tackle such challenges, governments and international organizations around the world are increasingly interested in issuing smart sanctions lists, which contain the names of entities involved in such criminal acts~\cite{Nephew2017, Zarate2013}.
	
    However, these lists vary quite substantially, both in the set of entities being banned and the timing of their inclusion, even when the target category that they aim to ban is the same.  This difference stems from several sources.  For instance, some institutions might be vigorous in collecting intelligence that influences others to follow, whereas other lists might be ignored because of a different understanding of the problem.  Some authoritative institutions might have a higher standard for banning an entity, while others might take actions without much undeniable evidence.  It is also possible that an institution is merely copying the entities added to another, prestigious list to project a sense of international cooperation without actively collecting intelligence themselves \cite{Furukawa2017}.  Thus, by analyzing the similarities and differences among such lists, we can shed some light on the influence network that governs the institutions issuing these lists.  The understanding of such an influence network could enrich our knowledge of the global political structure.

	In this paper, we describe the application of network analysis to the influence network that governs the major institutions issuing smart sanctions lists.  We use a dataset that includes over 1,700 smart sanctions lists, mainly focused on banning global criminal activities, and perform a Helmholtz--Hodge decomposition on the network constructed from this dataset.  We show that this simple analysis readily provides meaningful results, which enables a better understanding of the global political process behind smart sanctions lists.

	Many empirical studies measuring the impact of economic sanctions have restricted their scope to state-level effects, focusing on the period in which a comprehensive sanction was enacted at the state level~\cite{Dreger2015}. Meanwhile, studies focused on economic sanctions targeting specific entities such as firms, individuals, and vessels (that is, smart sanctions) are relatively new and less well understood, and studies using firm-level data are scarce~\cite{Ahn2019}. Of the few such studies to date,~\cite{Stone2016} undertook an event study and measured the impact of sanctions-related news items on the stock market. To the best of our knowledge, no study has yet focused on analyzing the influence structure among institutions issuing smart sanctions lists by constructing an influence network from the smart sanctions lists they have released.  The current paper opens the door to applying and developing a new network algorithm to study the foreign power structure in the global society, which would significantly affect political sciences.
	
    The rest of the paper is organized as follows.  In the next section, we review the dataset used in this paper.  We describe how we construct the influence network from a set of smart sanctions lists.  In Section 3, we briefly explain the Helmholtz--Hodge decomposition technique used in this paper.  In Section 4, we summarize our results.  The final section concludes the paper.

	%With that in mind, we perform network analysis on the influence network that governs the major institutions issuing smart sanctions list.  We would use a dataset that includes over 1,700 smart sanctions list that is mainly focused on banning global criminal activities.  We first give a basic description of the dataset and show how to construct the influence network from these sanctions lists.  We then perform a Helmholtz--Hodge decomposition on this network to understand hierarchical lead-lag influence structure of the network.  We show that this simple analysis already provides sensible results opening the door for a better understanding of the global political process of smart sanctions lists.
	
	\begin{table}[!htp]
		\centering
		\begin{tabular}{llr}
			\toprule
			Rank &                   Institution &  Count \\
			\midrule
			1 &                        European Union &    925 \\
			2 &  United States &    160 \\
			3 &                     Japan &    151 \\
			4 &                    Canada &     53 \\
			5 &            United Nations &     39 \\
			6 &               Switzerland &     39 \\
			7 &            United Kingdom &     35 \\
			8 &                 Australia &     19 \\
			9 &                 Singapore &     17 \\
			10 &                    Brazil &     15 \\
			%11 & South Korea	& 13 \\
			%12 & India	& 12 \\
			%13 & Germany &	11 \\
			%14 & Hong Kong & 	11 \\
			%15 & Malaysia &	11 \\
			%16 & Russia	& 9 \\
			%17 & Taiwan &	8 \\
			%18 & China &	8 \\
			%19 & Turkey &	7 \\
			%20 & Philippines &	7\\
			%21 & New Zealand & 7\\
			%22 & Ukraine & 7\\
			%23 & Spain & 7\\
			%24 & Mexico & 6\\
			%25 & Netherlands & 5\\
			%26 & France	& 5\\
			%27 & INTERPOL &	4\\
			%28 & Poland &	4\\
			%29 & Isle of Man &	4\\
			%30 & Vietnam &	4\\
			\bottomrule
		\end{tabular}
		\caption{
			Ranking of institutions by how many smart sanctions lists they have issued.
		}
		\label{table:institutions}
	\end{table}%[tbhp]	

	\section{Data}
	
	% Smart sanctions lists are available online for public use.  However, named entity recognition of banned targets might be a difficult task due to the variety of ways each institution handles their lists.  Thus 
    Smart sanctions lists are available online for public use.  However, named entity recognition of banned targets, such as companies and individuals, can be a difficult task because of the variety of ways each institution handles their lists.  Thus, we resort to information provided from professional sources.\footnote{Currently, we are recollecting the data for open use.}  We use the smart sanctions list data included in the Dow Jones Adverse Media Entities, State-Owned Companies, and the Watchlist datasets.  The Dow Jones datasets contain approximately 1,700 smart sanctions lists from 2001 to the present.  The purpose behind these smart sanctions are to curb illegal activities such as money laundering, drug use, fraud, organized crime, human trafficking, banned air carriers, and terrorist activity.

	A total of 85 institutions (such as countries and international organizations) have issued sanctions lists in our dataset.   The top ten institutions, in terms of the total number of sanctions lists issued, are provided in Table 1.  We can see that the majority are from countries worldwide, but international organizations such as the EU and the United Nations are also included.  The number of smart sanctions lists issued by each institution varies quite substantially.  The average number of lists issued is 16.7, and the standard deviation is 92.7, which confirms this insight.

    Each entry in a sanctions list comprises the name of the entity sanctioned and the date of their inclusion, and we can build two types of influence networks from this dataset.  One is the influence network at the smart sanctions list level, which treats each smart sanctions list as a node.  The network is constructed as follows. For each pair of smart sanctions lists, if list B includes an entity that is the same as an entity added earlier to list A, we add a weight (that is, 1) to the edge from A to B.  A pair of smart sanctions lists have no connecting edges if there are no common target entities in the lists.  We ignore cases in which two lists include entities precisely the same date because the direction of influence is not clear.

\begin{figure}[!htp]
		\begin{subfigure}{.62\textwidth}
			\centering
			\includegraphics[width=1.0\linewidth]{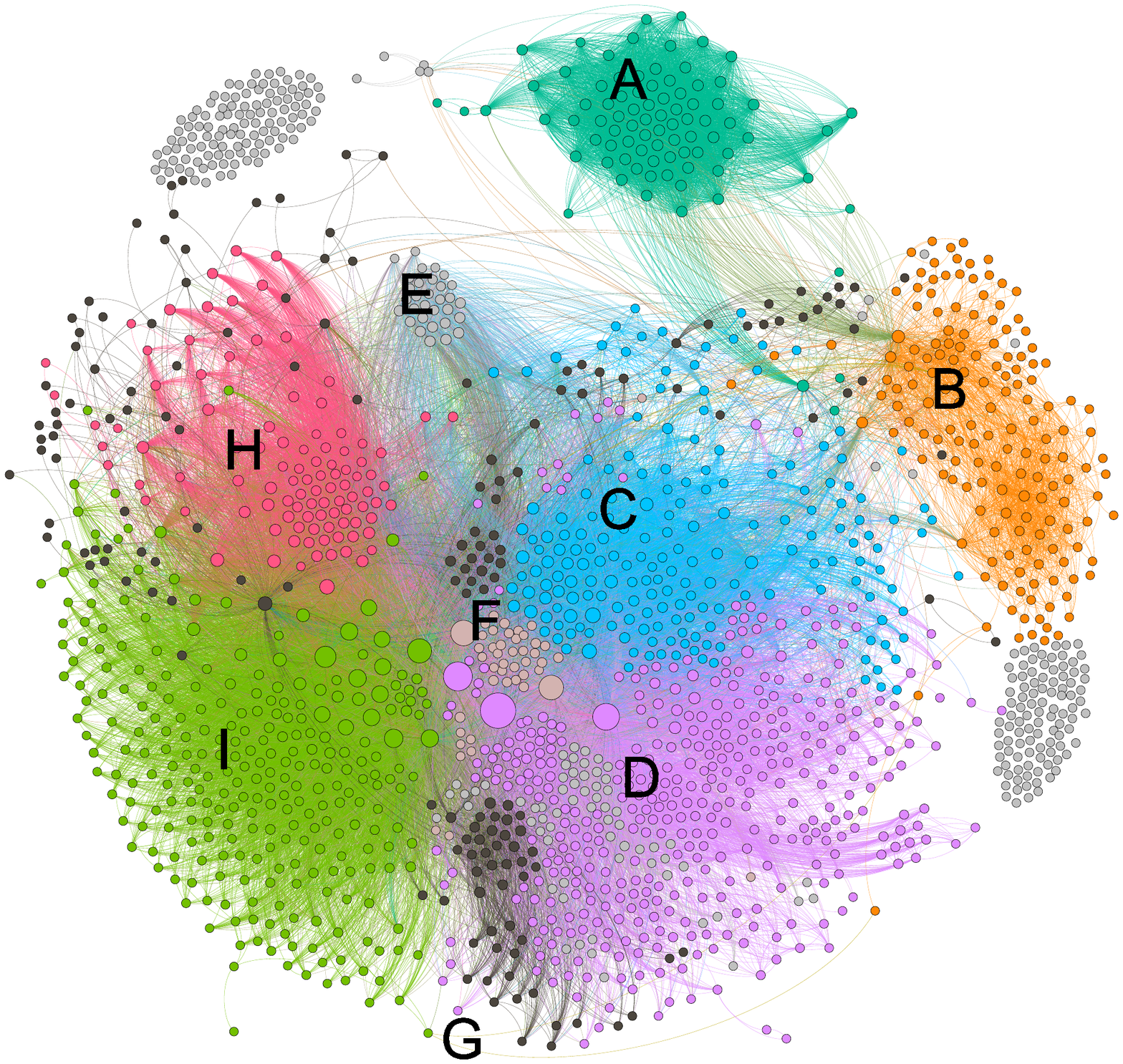}
			\caption{Influence network at the level of smart sanctions lists}
			\label{fig:list:network}
		\end{subfigure}	
		\begin{subtable}{.3\textwidth}
			\centering
			\resizebox{1.0\textwidth}{!}{
				\begin{tabular}{|p{2cm}|p{3cm}|}
					\toprule
					Community & Characteristic\\
					\midrule
					A & Issued by Japanese bureaucracy \\
					B & Against financial crimes  \\
					C & Against Iran and North Korea \\
					D &	Against Africa \\
					E &	Embargos \\
					F &	Against Burma \\
					G &	Against Libya \\
					H & Against terrorism in general\\
					I & Against Al-Qaeda \\
					\bottomrule
			\end{tabular}}
			\caption{Characteristics of each community}
			\label{table:list:community}
		\end{subtable}
		\caption{Analysis of the community structure of the influence network at the level of smart sanctions lists}
	\end{figure}

	We show the result in Fig.~\ref{fig:list:network}. The color of the nodes indicates the major communities found by standard modularity minimizing algorithms~\cite{Blondel2008, Lambiotte2008}.  We see many isolated nodes (that is, lists) located at the top of community H and below community B.  These lists are mainly domestic wanted lists issued by countries that are of little interest to other institutions.  Excluding these lists, the algorithms identified nine major communities. Table~\ref{table:list:community} summarizes the characteristics of each community.  We observe that smart sanctions lists targeting terrorism are generally located on the left (in community H) close to community I, which targets Al-Qaeda. On the bottom of the network, there is a community that targets Libya (community G), and right next to it, we have a community that deals with Africa in general.  Smart sanctions lists dealing with domestic Japanese issues have, in many cases, nothing in common with foreign lists, which explains why community A is isolated.  However, we can also see several edges between communities A and B, where B deals with financial crimes.  We use these distinctive communities (the categories of smart sanctions lists) in the following sections.
	
	\begin{figure}[!htp]
		\centering
		\includegraphics[width=1.0\linewidth]{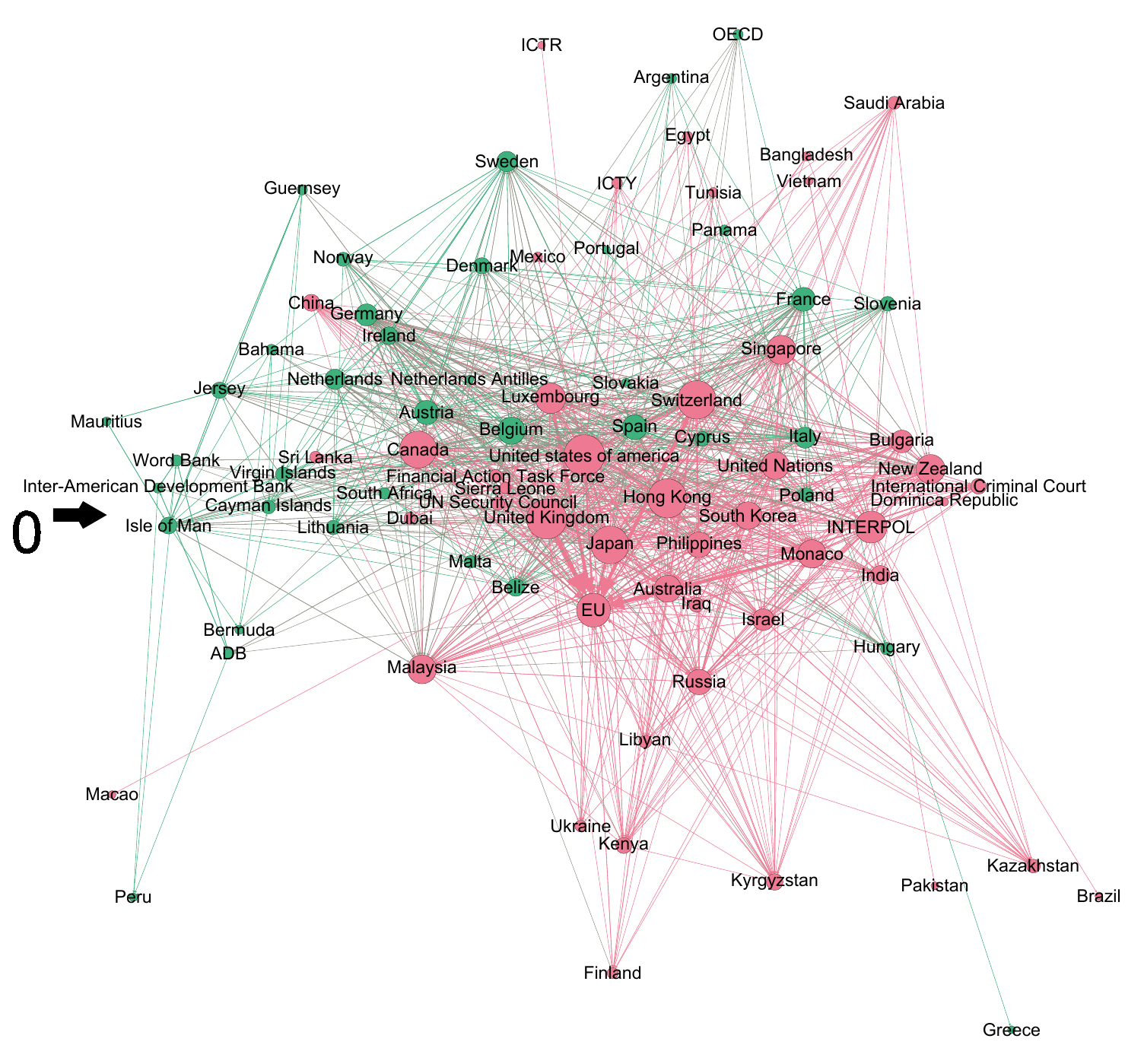}
		\caption{Influence network at institution level using all smart sanctions lists depicted in Fig.~\ref{fig:list:network}.}
		\label{fig:network}
	\end{figure}%

	We can also create an influence network at the country level, and this network is used to identify the hierarchical influence structure among the institutions behind the smart sanctions.  The steps taken are quite similar to those used to create the influence network at the smart sanctions list level.  We first treat each institution that has issued a list as a node.  For each pair of institutions, if institution B included the same entity on their list at a later time than institution A, we add a weight (that is, 1) to the edges between A and B.  A pair of institutions have no edges if there are no common entities on the lists that they have issued.  We ignore cases when two institutions add an entity on precisely the same date, as the direction of influence is not clear.  This procedure produces a weighted directed network of institutions, as shown in Fig.~\ref{fig:network}.  The network can be decomposed into two communities, distinguished by two colors, using standard modularity maximization techniques~\cite{Blondel2008, Lambiotte2008}. Note that the position on the y-axis reflects the hierarchical position (that is, the Helmholtz--Hodge potential) defined in the next section.

	\section{Helmholtz--Hodge Decomposition}
	
	\begin{figure}[htp]
	\centering
	\includegraphics[width=0.52\linewidth]{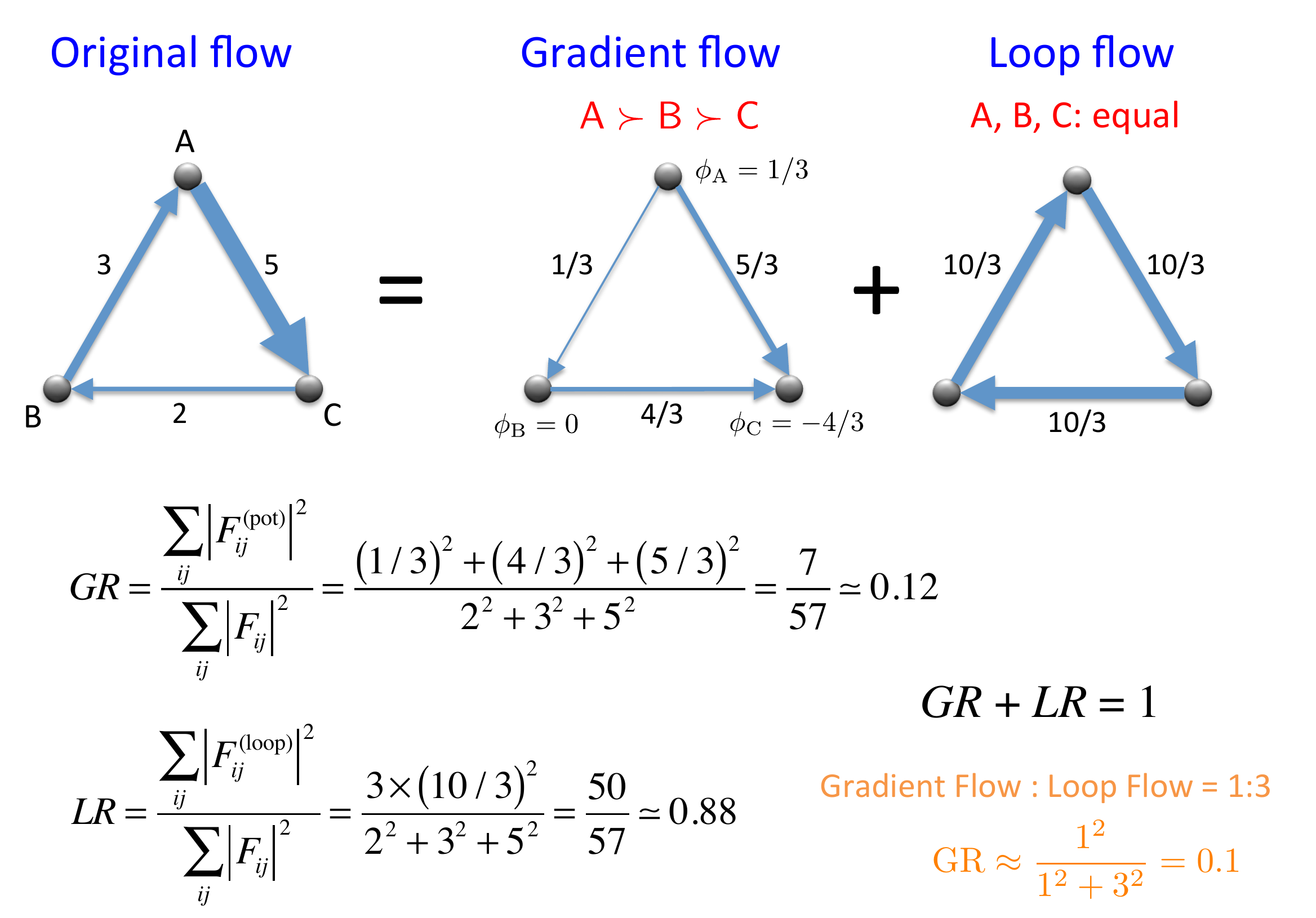}
	\caption{Illustration of gradient and loop ratios.}
	\label{fig:ratio}
	\end{figure}
	
	The flow $F_{ij}$ running from node $i$ to node $j$ in a directed network can be decomposed into     
	
	\begin{equation}
	F_{ij} = F_{ij}^{p} + F_{ij}^{c},
	\end{equation}
	
	\noindent where $F_{ij}^{p}$ denotes the gradient flow and $F_{ij}^{c}$ denotes the circular flow (see Fig.~\ref{fig:ratio} for a visual illustration).  Circular flow $F_{ij}^{c}$ corresponds to the feedback loops that are inherent in such networks.   Gradient flow $F_{ij}^{p}$ can be understood as the hierarchical component of the network, where information flows from nodes with higher potentials to nodes with lower ones.  Mathematically, this can be written as

	\begin{equation}
	F_{ij}^{p} = w_{ij}(\phi_{i}-\phi_{j}),
	\end{equation}
	
	\noindent where $w_{ij}$ is the weight of the edges between nodes $i$, and $j$ and $\phi_{i}$ denotes the Helmholtz--Hodge potential associated with node $i$.  The Helmholtz--Hodge potential of a node reflects its hierarchical position in its flow structure, which neglects the effect from the feedback mechanism.  The potential $\phi_{i}$ for every node can be easily determined by minimizing the overall squared difference between the actual flow and the gradient flow (see \cite{Jiang2011, Kichikawa2019} for more details).
	
	%\begin{equation}
	%F_{ij}^{p} = w_{ij}(\phi_{i}-\phi_{j})
	%\end{equation}
	
	%\noindent where $w_{ij}$ is a weight for edges between nodes $i$ and $j$ and $\phi_{i}$ denotes the Helmholtz-Hodge potential associated with node $i$.  Helmholtz-Hodge potential of nodes reflects their hiererachical postision in its flow stucture ignoring the effect coming from feedback mechanism.  The determination of the potential $\phi_{i}$ for every node could be easily determined by minizing the squared difference between the actual flow and the gradient flow. See \cite{Jiang2011} for more detail.% as in 
	
	%\begin{equation}
	%L=\frac{1}{2} \Sigma_{i<j}w_{ij}^{-1}(F_{ij}-F_{ij}^{p})^{2}
	%\end{equation}

	%\begin{equation}
	%F_{ij}^{p} = w_{ij}(\phi_{i}-\phi_{j})
	%\end{equation}
	
	%\noindent where $w_{ij}$ is a weight for edges between nodes $i$ and $j$ and $\phi_{i}$ denotes the Helmholtz-Hodge potential associated with node $i$.  Helmholtz-Hodge potential of nodes reflects their hiererachical postision in its flow stucture ignoring the effect coming from feedback mechanism.  The determination of the potential $\phi_{i}$ for every node could be easily determined by minizing the squared difference between the actual flow and the gradient flow. See \cite{Jiang2011} for more detail.% as in 
	
	%\begin{equation}
	%L=\frac{1}{2} \Sigma_{i<j}w_{ij}^{-1}(F_{ij}-F_{ij}^{p})^{2}
	%\end{equation}
	
	\section{Results}
	\subsection{Results using all lists}

	\begin{figure}
		\begin{subfigure}{.55\textwidth}
			\centering
			\includegraphics[width=.95\linewidth]{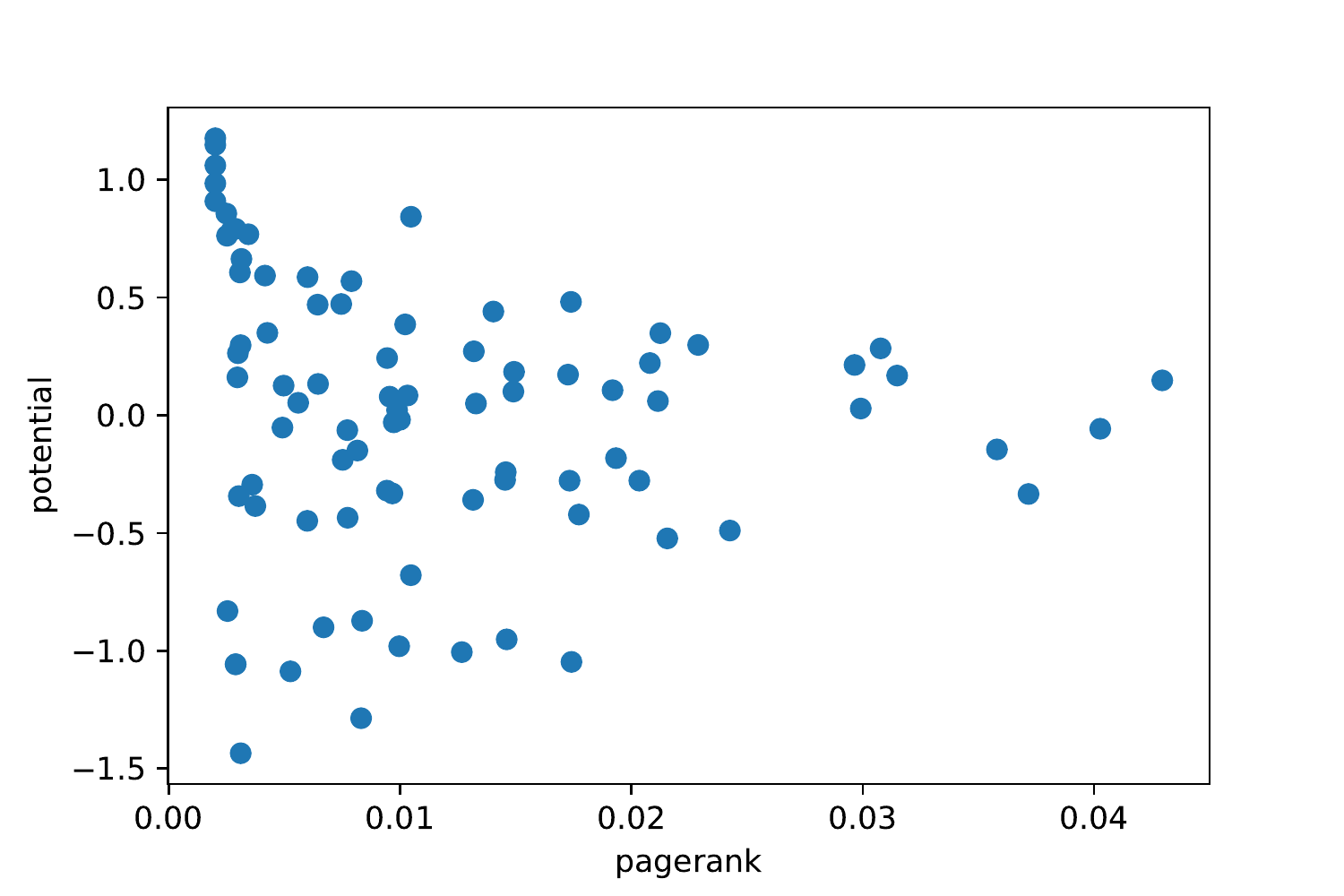}
			\caption{Scatterplot comparing the page rank value and Helmholtz--Hodge potential of each node.}
			\label{fig:pagerank}
		\end{subfigure}	
		\begin{subtable}{.45\textwidth}
			\centering
			\resizebox{1.0\textwidth}{!}{
				\begin{tabular}{llr}
					\toprule
					Rank &                   Institution &  Potential \\
					\midrule
					1 & OECD	& 1.176 \\
					2 & ICTR &	1.148 \\
					%3 & Argentina &	1.06\\
					%4 &	Saudi Arabia &	0.985\\
					%5 &	Egypt &	0.909\\
					%6 &	Bangladesh &	0.857\\
					%7 &	Sweden &	0.843\\
					%8 &	Vietnam &	0.792\\
					%9 &	ICTY &	0.787\\
					%10 &	Guernsey &	0.769\\
					%11 &	Tunisia &	0.762\\
					%12 &	Panama &	0.664\\
					%13 &	Portugal &	0.607\\
					%14 &	Mexico &	0.593\\
					%15 &	Norway &	0.587\\
					%16 &	Denmark &	0.570\\
					17 &	\textbf{France} &	0.482\\
					%18 &	China &	0.473\\
					%19 &	Slovenia &	0.470\\
					20 &	\textbf{Germany} &	0.441\\
					%21 &	Ireland &	0.386\\
					%22 &	Bahama &	0.350\\
					%23 &	Singapore &	0.349\\
					%24 &	United Nations &	0.299\\
					%25 &	Sierra Leone &	0.298\\
					26 &	\textbf{United Kingdom} &	0.284\\
					%27 &	Netherlands &	0.272\\
					%28 &	Slovakia &	0.264\\
					%29 &	Jersey &	0.243\\
					%30 &	Luxembourg &	0.222\\
					%31 &	Switzerland &	0.214\\
					%32 &	Austria &	0.185\\
					%33 &	Spain &	0.173\\
					%34 &	Hong Kong &	0.169\\
					%35 &	Mauritius &	0.161\\
					36 &	\textbf{United States of America} &	0.148\\
					%37 &	Word Bank &	0.133\\
					%38 &	South Africa &	0.126\\
					%39 &	New Zealand &	0.107\\
					40 &	\textbf{Italy} &	0.101\\
					%41 &	Cyprus &	0.084\\
					%42 &	Isle of Man &	0.079\\
					%43 &	Belgium	 & 0.060\\
					%44 &	Inter-American Development Bank &	0.053\\
					%45 &	Bulgaria &	0.050\\
					46 &	\textbf{Canada} &	0.029\\
					%47 &	Cayman Islands &	0.022\\
					%48 &	Poland &	-0.019\\
					%49 &	Virgin Islands &	-0.030\\
					%50 &	Dubai &	-0.052\\
					%51 &	INTERPOL &	-0.057\\
					%52 & Lithuania &	-0.063\\
					53 & \textbf{Japan} &	-0.145\\
					%54 & Financial Action Task Force &	-0.150\\
					%55 & Monaco	 & -0.182\\
					%56 & Malta &	-0.189\\
					%57 & India &	-0.242\\
					%58 & Belize &	-0.274\\
					%59 & Philippines &	-0.277\\
					%60 & Australia &	-0.278\\
					%61 & Dominica Republic &	-0.295\\
					%62 & Iraq &	-0.320\\
					%63 & International Criminal Court & -0.332\\
					%64 & EU	 & -0.334\\
					%65 & Netherlands Antilles &	-0.343\\
					%66 & Israel &	-0.359\\
					%67 & Bermuda &	-0.385\\
					%68 & South Korea &	-0.421\\
					%69 & Hungary &	-0.435\\
					%70 & ADB &	-0.448\\
					%71 & Malaysia &	-0.489\\
					%72 & Russia &	-0.523\\
					%73 & Libyan &	-0.679\\
					%74 & Macao &	-0.831\\
					%75 & Sri Lanka &	-0.872\\
					%76 & Ukraine &	-0.900\\
					%77 & Kenya &	-0.951\\
					%78 & UN Security Council &	-0.981\\
					%79 & Kazakhstan &	-1.006\\
					%80 & Kyrgyzstan &	-1.047\\
					%81 & Pakistan &	-1.057\\
					%82 & Brazil	 & -1.057\\
					%83 & Peru	& -1.087\\
					%84 & Finland &	-1.286\\
					%85 & Greece	& -1.435\\
					\bottomrule
			\end{tabular}}
			\caption{Estimated Helmholtz--Hodge potentials. G7 countries are denoted in bold font. }
			\label{table:potential}
		\end{subtable}
	\end{figure}%

    In Fig.~\ref{fig:pagerank}, we show a scatterplot comparing the estimated potentials and the page rank value of each node~\cite{Brin1998}.  We confirm that the Helmholtz--Hodge potential estimated by the Helmholtz--Hodge decomposition reveals information that is independent of page rank value.

	\begin{figure}[!htp]
	\centering
	\includegraphics[width=.6\linewidth]{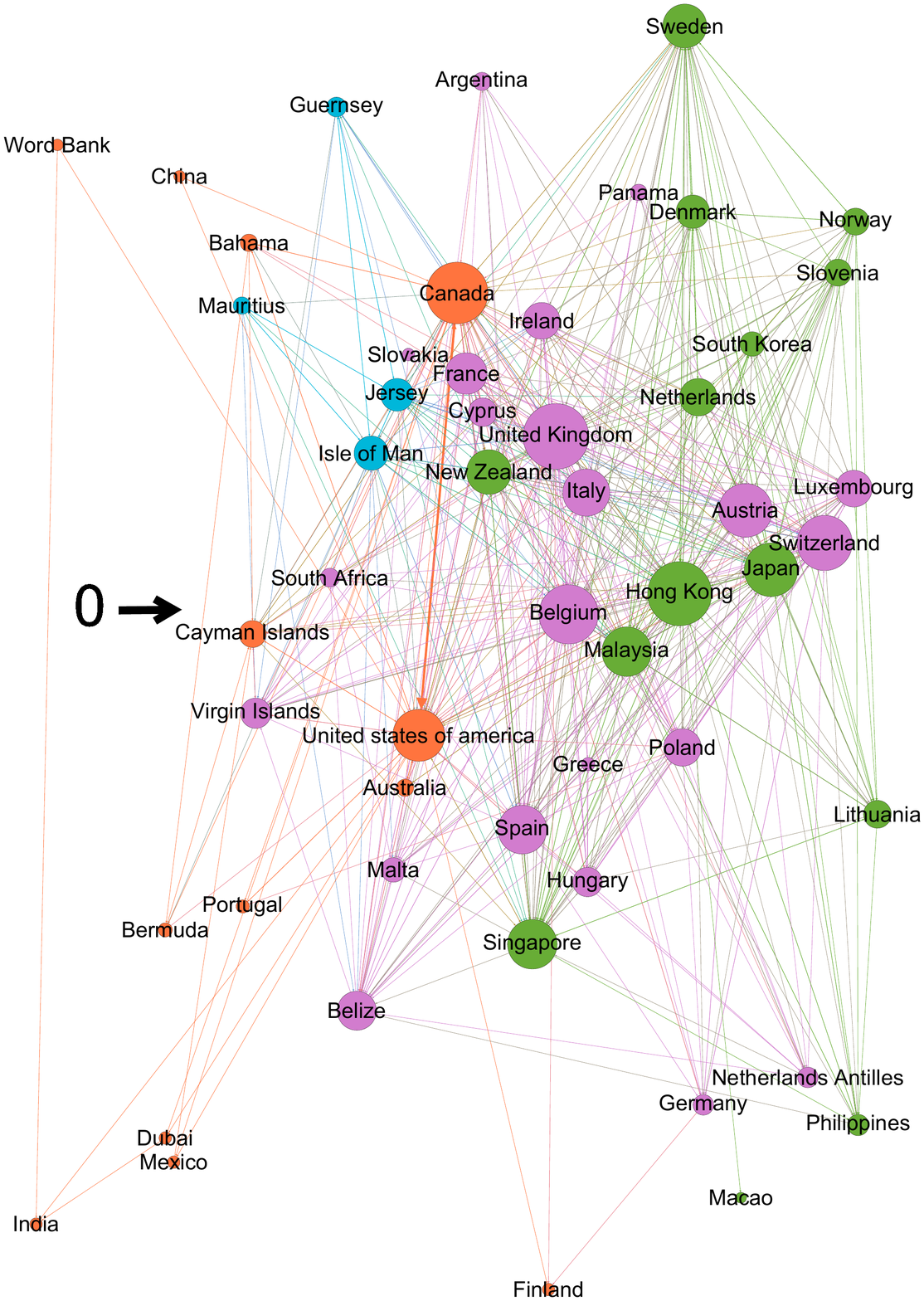}
	\caption{Influence network at institution level using the smart sanctions lists grouped as financial crimes in Fig.~\ref{fig:list:network}.}
	\label{fig:14}
\end{figure}	

    Table~\ref{table:potential} shows a subset of the estimated Helmholtz--Hodge potential for the 85 institutions analyzed in this paper (the potential of the rest of the institutions is indicated by their position on the y-axis in Fig. 2\footnote{The actual position were slightly adjusted so that the nodes do not overlap.}).  The arrow shows the location of where the y-axis being 0.  The G7 countries (Canada, France, Germany, Italy, Japan, the United Kingdom, and the United States of America) are shown in bold font.  We can see that the OECD and the International Criminal Tribunal for Rwanda (ICTR), which are primarily focused on specific issues, top the list, which shows that they are less influenced by the sanctions lists issued by other institutions.  This observation is quite intuitive because the smart sanctions issued by these institutions are restricted to focusing on specific issues.  However, an analysis using all the smart sanctions lists is somewhat vague, locating a large proportion of the institutions in the middle.  To look deeper into the structure, we must divide the smart sanctions lists into categories.

	%Among the G7 countries, Japan scores the lowest.  This is to be expected, because Japan is relatively inactive when it comes to gathering information outside of Japan and merely follows lists created by international organizations~\cite{Furukawa2017}.  The fact that the UN Security Council and EU are quite low on the list is also notable, as these two organizations require a higher standard of banning compared to lists issued at a country level.  To summarize, our analysis provides meaningful insights into the hierarchical influences underlying smart sanctions lists.

	\subsection{Restricting lists to categories}
	
	    	\begin{table}[!htp]
		\centering
		\begin{tabular}{llr}
			\toprule
			Category & Gradient & Loop \\
			\midrule
			Libya & 0.865 &  0.135 \\
			Africa & 0.886 & 0.114 \\
			Burma & 0.883 &  0.117 \\
			Terrorism in general & 0.98 &     0.02 \\
			Al-Qaeda &  0.79 &  0.21 \\
			Iran-North Korea &  0.826 &  0.174 \\
			\bottomrule
		\end{tabular}
		\caption{
			Gradient and loop ratio for each category.
		}
		\label{table:loop}
	\end{table}%[tbhp]	

    In this section, we provide results obtained by restricting the set of lists used to derive the influence network.  We use seven out of the nine categories found, as depicted in Fig. 1, excluding categories A (that is, issued by Japanese bureaucracy) and E (embargoes) because only a few countries were involved in the country-level network created from these smart sanction lists.  The gradient and loop ratio are relative measures quantifying to what extent the flow structure is hierarchical and circular respectively (i.e., Fig.~\ref{fig:ratio}).  The two ratios are thereby complementary to each other so that the sum of them is always normalized to 1.  Table~\ref{table:loop} summarizes the gradient and loop ratios of all the networks.    We observe that the loop ratio ranges from 0.11 to 0.21 for almost all the network, which indicates that a non--negligible amount of information loops through the network.  However, for smart sanctions against terrorism in general, the loop ratio is close to 0.0, which means that there is almost no loop structure.  Hierarchical components dominate the network, and information flows to a hub, as shown in Fig. 6(b).  In all cases, the average value of the Helmholtz-Hodge potential is set to 0.0.  Thus, if the potential of a node is positive, we can conclude that it is located in the upper stream side, and vice versa.

    In Fig.~\ref{fig:14}, we show the result of our Helmholtz--Hodge analysis for the financial crime category (that is, B in Fig.~1).  As in Fig.~2, the y-axis position is determined using the Helmholtz--Hodge potential and the x-position is determined using the method of ~\cite{Noack2007}.  In this influence network, we see that there are many players involved, which creates four different communities (North America, Europe, island nations in the Indian Ocean, and other significant countries) where each community focuses on different sets of targeted entities.  Moreover, the North American community and the island nation's community are located in distinct positions on the left, which reflects the similarity between these communities.

\begin{figure}[!htp]
	\begin{subfigure}{.5\textwidth}
		\centering
		\includegraphics[width=1.0\linewidth]{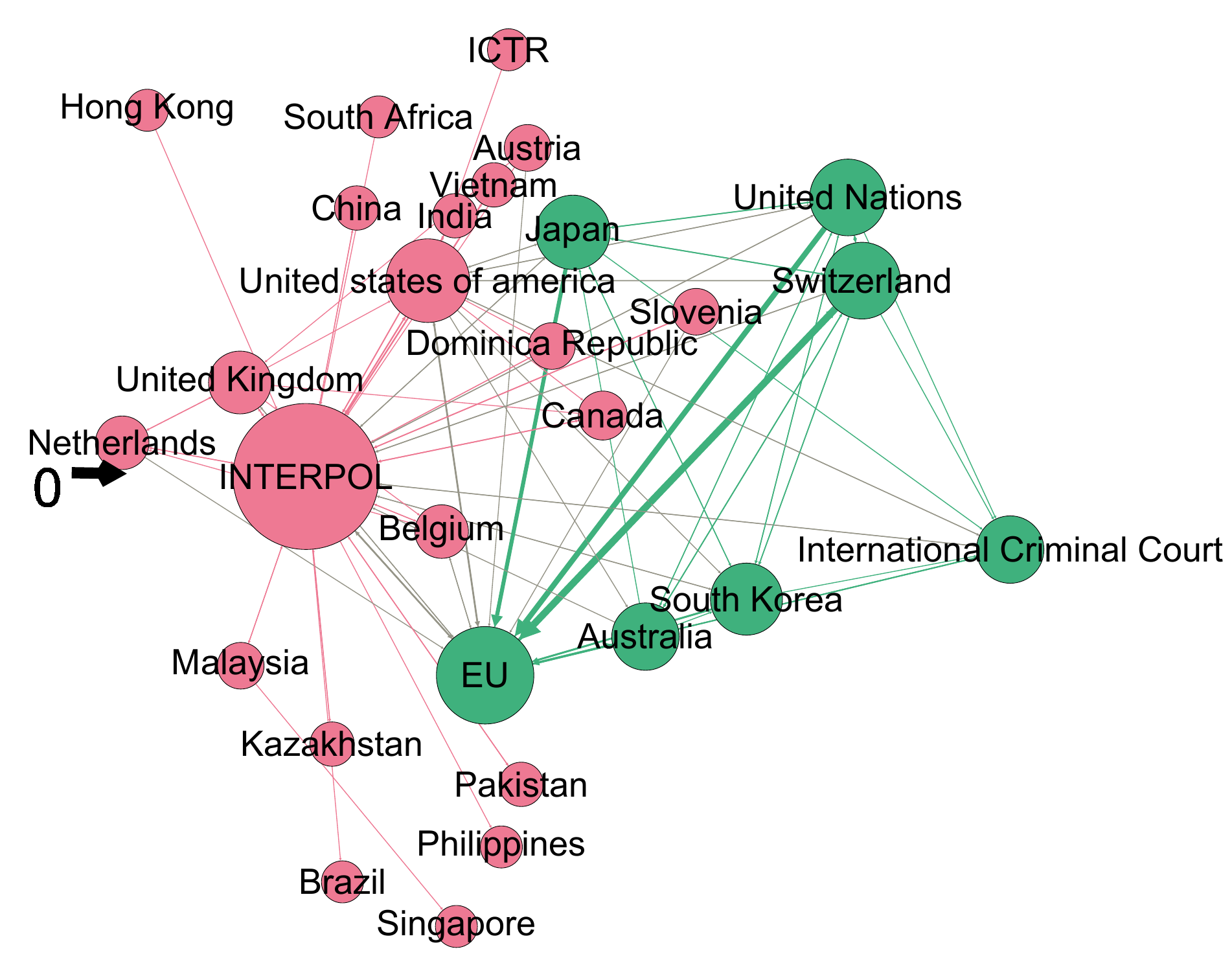}
		\caption{Influence network at the institution level using smart sanctions lists grouped as against Libya in Fig.~\ref{fig:list:network}.}
		\label{fig:34}
	\end{subfigure}	
	\begin{subfigure}{.5\textwidth}
		\centering
		\includegraphics[width=1.0\linewidth]{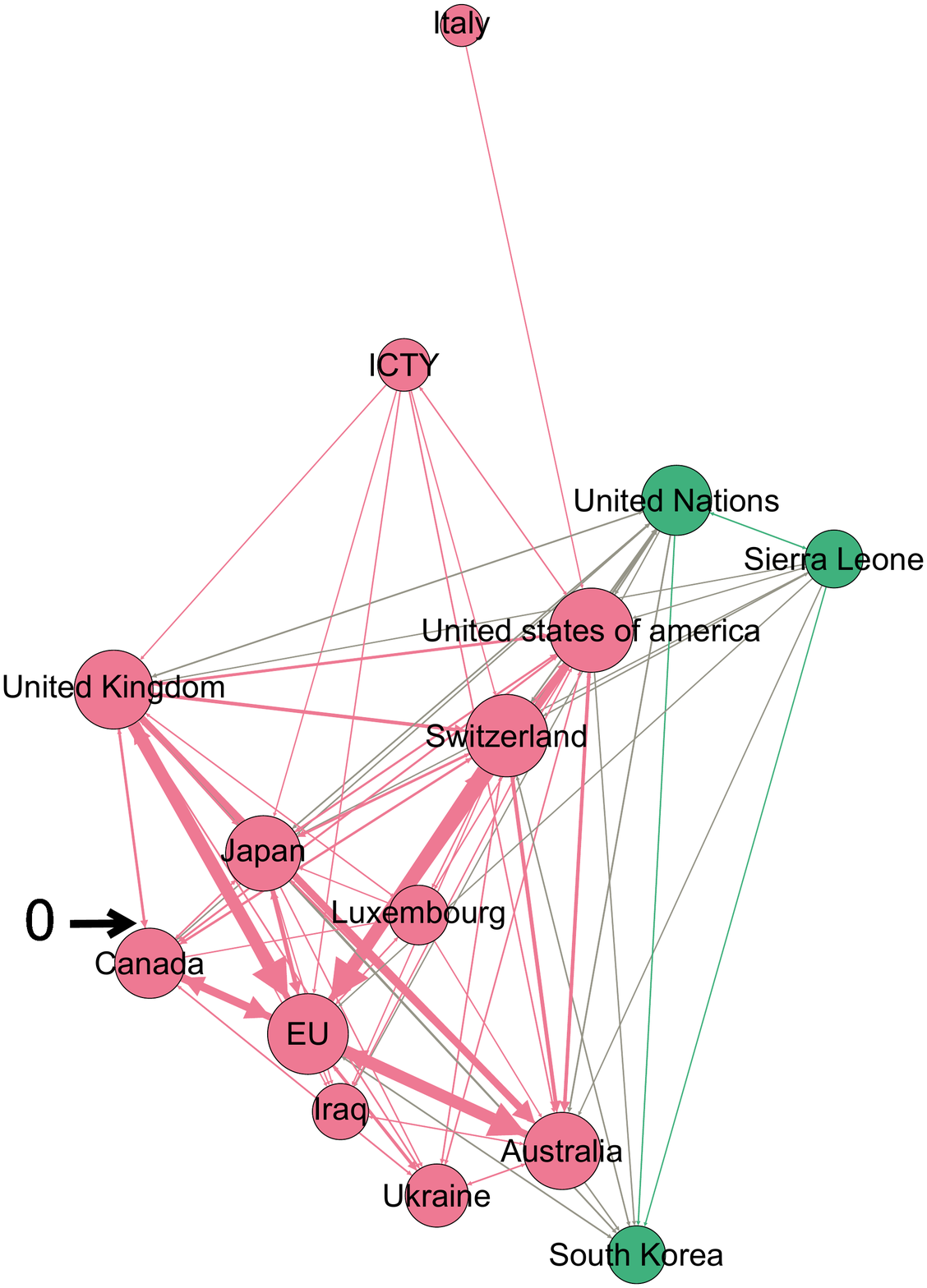}
		\caption{Influence network at the institution level using smart sanctions lists grouped as against Africa in Fig.~\ref{fig:list:network}.}
		\label{fig:37}
	\end{subfigure}	
	\caption{Analysis of the community structure of the influence network at the level of smart sanctions lists}
\end{figure}

	In Fig.~\ref{fig:34}, we show the influence network for the Libya category (G in Fig. 1).  In this example, we find two distinct communities.  One reflects countries and institutions that heavily influence the European Union (such as the United Nations, Switzerland, and Japan) and the other reflecting institutions that influence INTERPOL.  Figure~\ref{fig:37} shows the result for sanctions against Africa (D in Fig. 1).  Italy is at the far top of the hierarchy, but there is only one edge pointing to the United States.  Switzerland and the United Kingdom are key players that strongly influence the European Union, whereas Australia seems to be a follower of the European Union decisions.

\begin{figure}[!htp]
	\begin{subfigure}{.5\textwidth}
		\centering
		\includegraphics[width=1.0\linewidth]{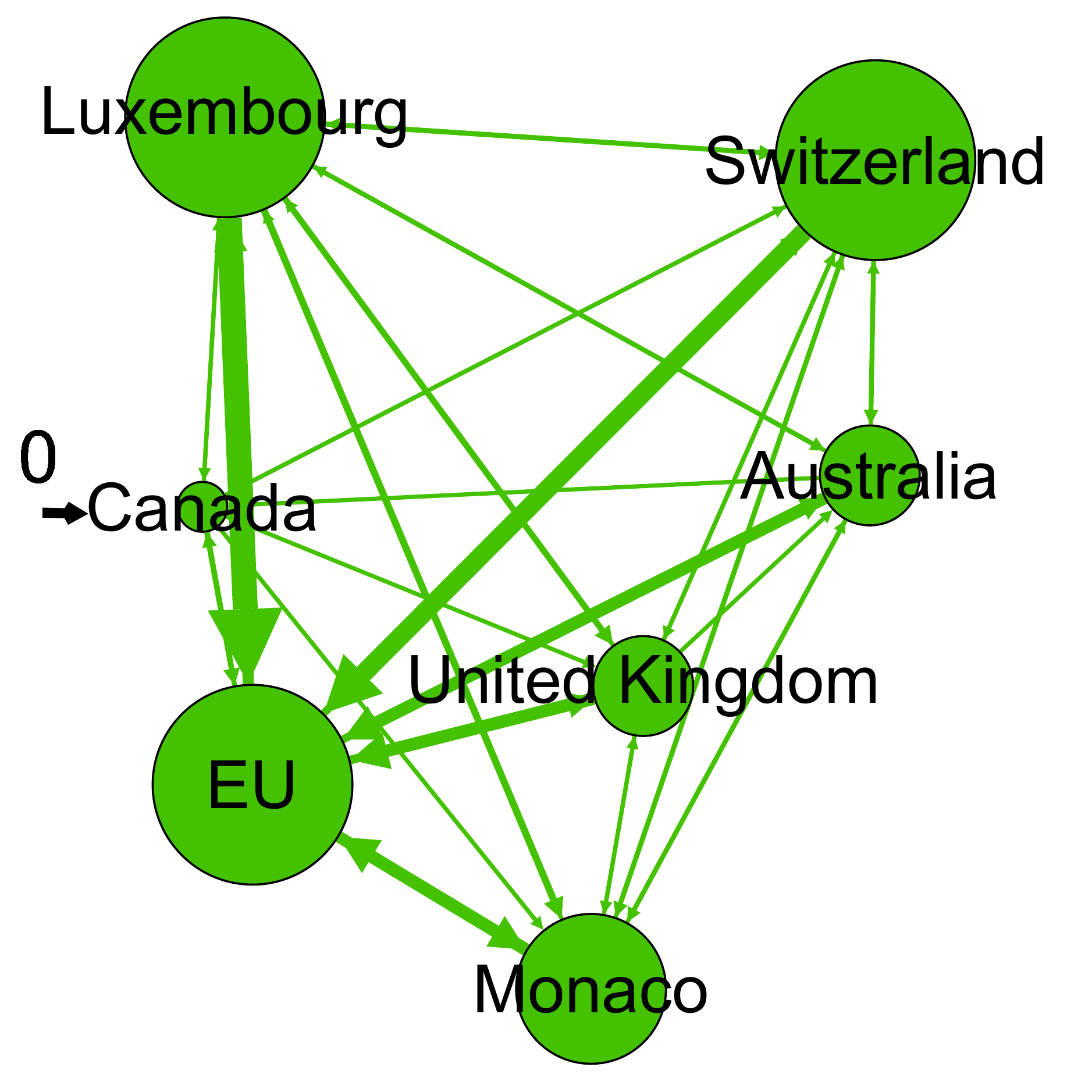}
		\caption{Influence network at the institution level using smart sanctions lists grouped as against Burma in Fig.~\ref{fig:list:network}.}
		\label{fig:92}
	\end{subfigure}	
	\begin{subfigure}{.5\textwidth}
		\centering
		\includegraphics[width=1.0\linewidth]{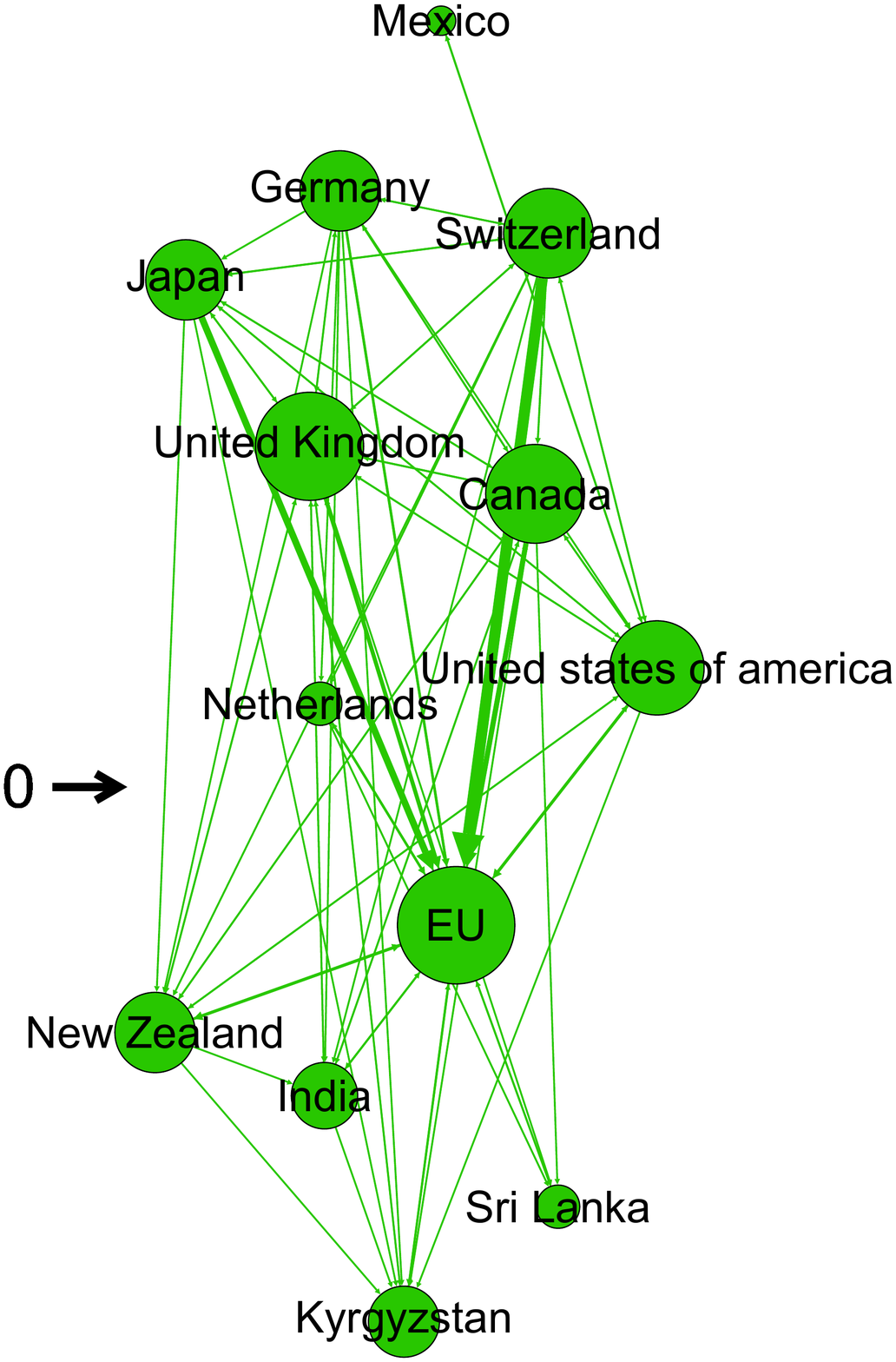}
		\caption{Influence network at the institution level using smart sanctions lists grouped as against terrorism in general in Fig.~\ref{fig:list:network}.}
		\label{fig:184}
	\end{subfigure}	
	\caption{Analysis of community structure of the influence network at the level of smart sanctions lists}
\end{figure}

    Figures~\ref{fig:92} and~\ref{fig:184} showcases in which there is only one community in the network.  Figure~\ref{fig:92} corresponds to smart sanctions against Burma (F in Fig. 1), and Fig.~\ref{fig:184} corresponds to terrorism in general.  In both cases, Switzerland is the key player that strongly affects the European Union's decision.  Luxembourg is another key player in the Burma case, but it is quite interesting that Switzerland and Luxembourg do not have much influence on each other.

\begin{figure}[!htp]
	\begin{subfigure}{.5\textwidth}
		\centering
		\includegraphics[width=1.0\linewidth]{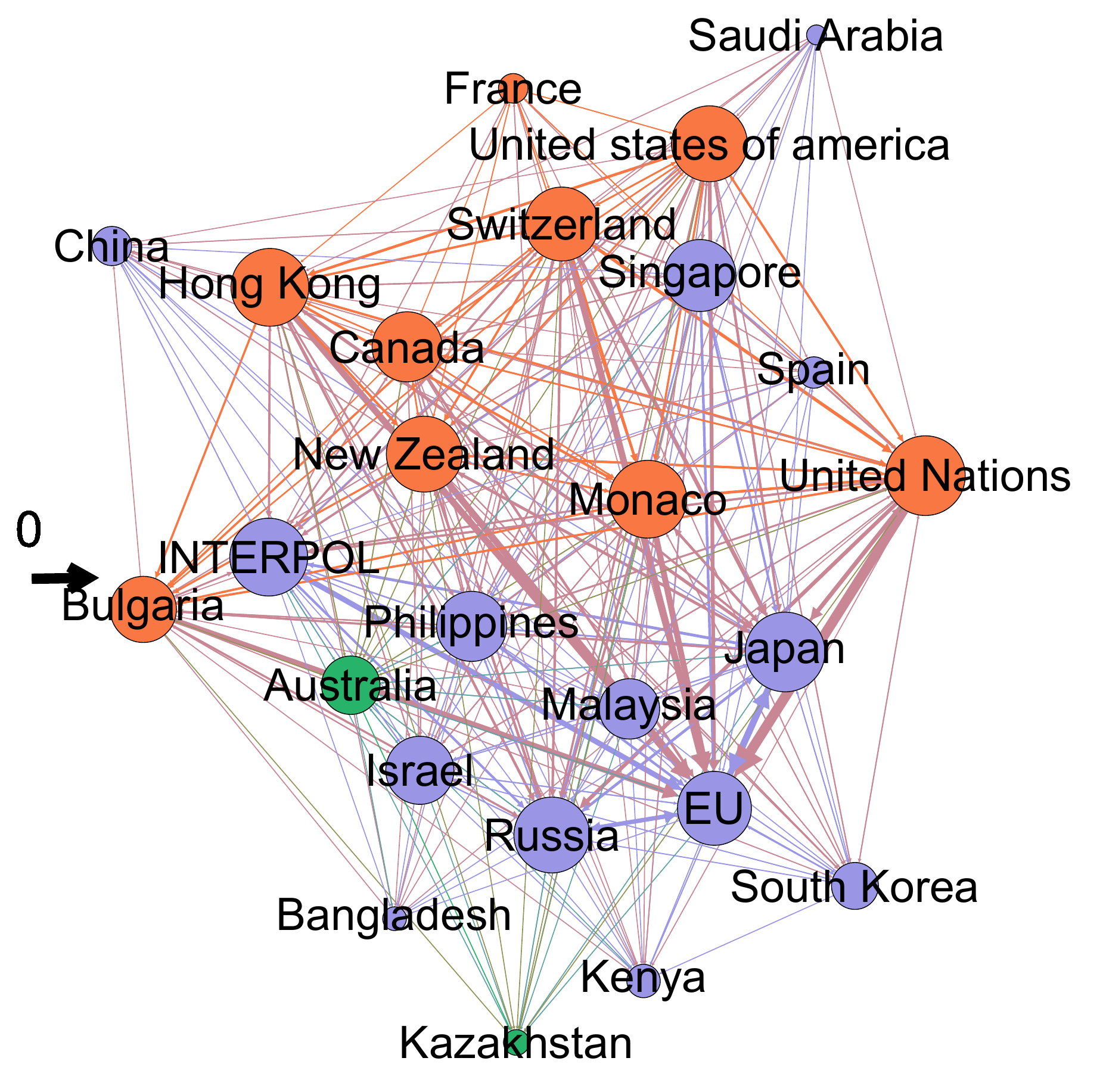}
		\caption{Influence network at the institution level using smart sanctions lists grouped as against Al-Qaeda in Fig.~\ref{fig:list:network}.}
		\label{fig:200}
	\end{subfigure}	
	\begin{subfigure}{.5\textwidth}
		\centering
		\includegraphics[width=1.0\linewidth]{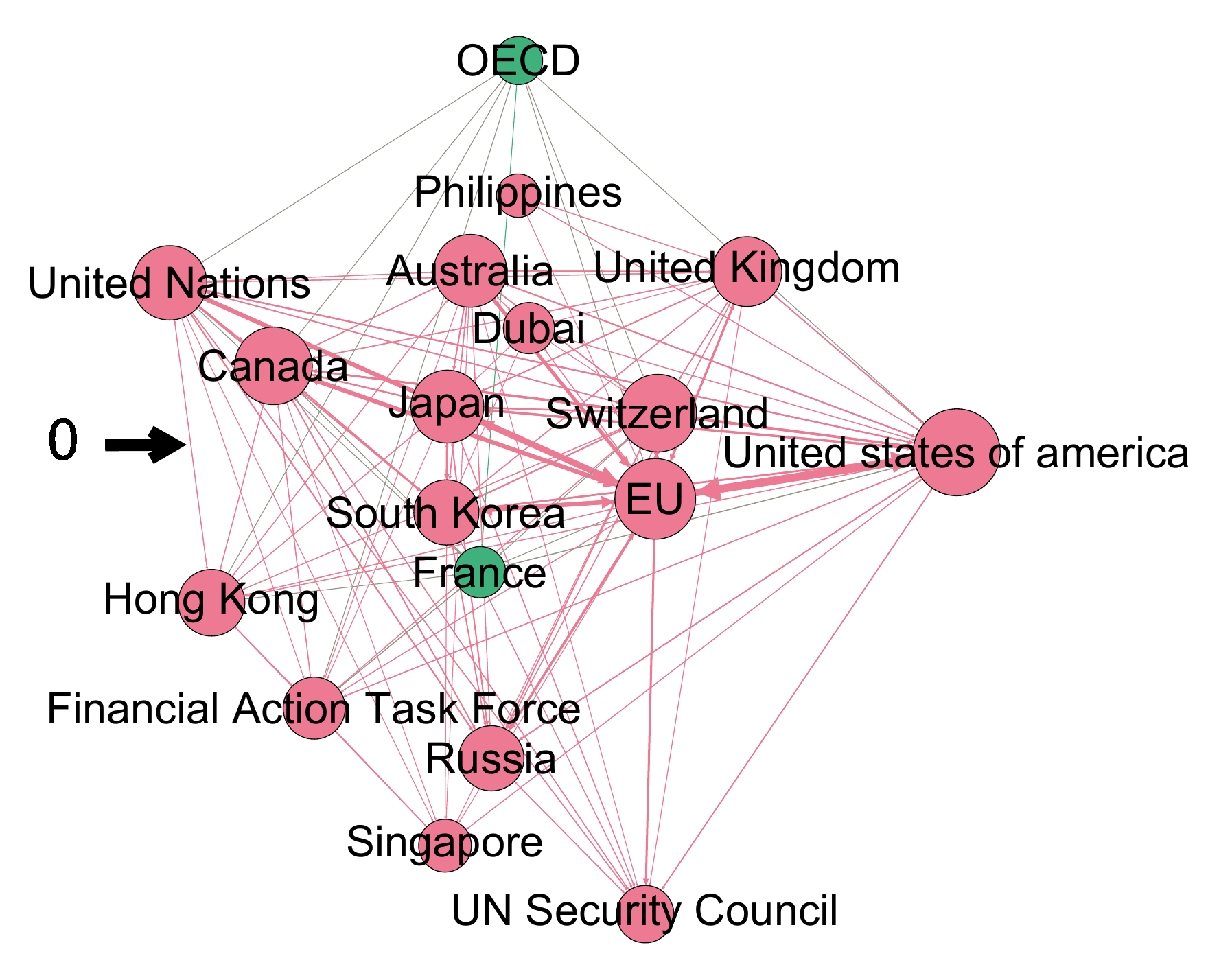}
		\caption{Influence network at the institution level using smart sanctions lists grouped as against Iran and North Korea in Fig.~\ref{fig:list:network}.}
		\label{fig:203}
	\end{subfigure}	
	\caption{Analysis of community structure of the influence network at the level of smart sanctions lists}
\end{figure}

    Figure~\ref{fig:200} shows the result for the Al-Qaeda case (I in Fig. 1).  This is a quite complex case in which there are three communities and many edges among the institutions, which makes the network quite dense.   We can see that the United Nations and the European Union are both working as hubs, gathering information from various countries. Still, the United Nations seems to be leading the European Union, as could be suspected by its position.  In Fig.~\ref{fig:203}, we show the case concerning the sanctions against Iran and North Korea.  In this case, we also see that the European Union acts as a hub, gathering information worldwide.   It is also noteworthy that the United Nations Security Council is apparently at the bottom of the hierarchy, which indicates the complex approval mechanism of this council~\cite{Furukawa2017}.

	Table~\ref{table:selected} summarizes the Helmholtz--Hodge potential for selected countries.  Several things are worth mentioning here. Switzerland appears in almost all categories, with positive values for each category.  This observation indicates that Switzerland is quite an active player when it comes to smart sanctions.  Australia, however, does appear in many categories but instead has a negative overall position (except for Iran-North Korea), which indicates that it is more of a follower.  The United States is notably in a high position for Al-Qaeda, Libya, Africa, and terrorism in general, but is located in the lower part for financial crimes.  Japan, conversely, is located in the lower part for Al-Qaeda but is an active player when it comes to terrorism in general. To summarize, our analysis provides meaningful insights into the hierarchical influences underlying smart sanctions lists.

	\begin{table}[!htp]
		\centering
		\resizebox{1.0\textwidth}{!}{
			\begin{tabular}{lrrrrrrr}
				\toprule
				Country & Financial Crimes & Libya & Africa	& Burma & Terrorism & Al-Qaeda	& Iran-North Korea \\
				\midrule
				Japan & 0.073 & 0.256 & 0.053 & - & 0.485 & -0.202 & 0.05 \\
				United States & -0.132 & 0.239 & 0.208 & - & 0.225 & 0.627 &  -0.032 \\
				China &	0.507 & 0.597 &	- & - & - & 0.486 & - \\
				Australia & -0.16 & -0.193 & -0.542 & 0.039	& - & -0.323 & 0.42\\
				United Kingdom & 0.158 & 0.031 & 0.207 & -0.046 & 0.446 &	- & 0.379\\
				Germany	& -0.392 & - & - & - & 0.576 & - & - \\
				France	& 0.198	& - & - & - & - & 0.883	& -0.21\\
				Italy & 0.114 & - & 1.208	& - & - & -\\
				Canada & 0.327 & 0.01 &	-0.105 & 0.0 & 0.356 & 0.243 & 0.122\\
				Switzerland &  0.087 & 0.186 & 0.191 & 0.294 & 0.54	 & 0.525 & 0.037\\
				\bottomrule
		\end{tabular}}
		\caption{
			Helmholtz-Hodge potential for selected countries.
		}
		\label{table:selected}
	\end{table}	
	
	\section{Conclusion}
	
	%In this paper, we made two contributions.  The first is that we showed how to construct an influence network governing smart sanctions at the country level from the smart sanctions lists they have issued.  We then showed that by performing a Helmholtz--Hodge decomposition of the influence network, we could shed some light on the influence network that governs the institutions issuing such lists.  We found that for some categories of smart sanctions lists, the influence network is governed almost entirely by a hierarchical structure, but for others there is a significant amount of loop flow meaning that countries and international institutions are actively influencing each other.  Our analysis opens the door to a better understanding of the hidden global political structure behind smart sanctions. 
	
    This paper makes two contributions. The first is that we showed how to construct an influence network governing smart sanctions at the country level from the smart sanctions lists each country has issued. We then showed that by performing a Helmholtz--Hodge decomposition of the influence network, we could shed some light on the influence network that governs the institutions issuing such lists.

	Specifically, we derived three interesting observations from our Helmholtz--Hodge analysis. First, we found that for smart sanctions lists against Iran and North Korea, while the United Kingdom, United Nations, and the United States are at the top of the hierarchy of influencing other major countries and international institutions, the United Nations Security Council is clearly at the bottom of the hierarchy, which indicates its complex approval mechanism.  Second, for the smart sanctions against Libya, the United Nations, ICTR, United States, and Japan are the key influencers in the upper stream of the hierarchy, and INTERPOL and the European Union are located in the middle of the hierarchy, acting as hubs aggregating information. Another intriguing example is the smart sanctions concerning financial crimes. In this example, there are four different communities (North America, Europe, island nations, and other countries), and each community focuses on different sets of targeted entities.
	
	Furthermore, for some of the smart sanctions list categories, the influence network is governed almost entirely by a hierarchical structure. Still, for others, there is a significant amount of loop flow, which indicates that countries and international institutions actively influence each other. Our simple and effective analysis enables a better understanding of the hidden global political structure behind smart sanctions.

	\section{Acknowledgements}
	
	We are grateful for Takaaki Ohnishi and Tsutomu Watanabe for helpful discussions. We thank Stuart Jenkinson, PhD, and Kimberly Moravec, PhD, from Edanz Group (www.edanzediting.com/ac) for editing a draft of this manuscript.
	
	\newpage
	
	\bibliographystyle{plain}
	\bibliography{Hisano_Bib}

\end{document}